\documentstyle[preprint,aps]{revtex}

\begin{document}

\draft

\preprint{$
\begin{array}{l}
\mbox{AMES-HET-97-06}\\[-3mm]
\mbox{May 1997}\\
\end{array}
$}

\title{$b \bar b$ Production on the $Z$ Resonance and Implications for LEP2}

\author{A. Datta$^a$, K.Whisnant$^a$, Bing-Lin Young$^a$, and X. Zhang$^b$}

\address{{$^a$Department of Physics and Astronomy, Iowa State University,
Ames, IA 50011, USA}\\
{$^b$Institute of High Energy
Physics, Beijing 100039, P.R. China}\\}
\maketitle

\begin{abstract}

In terms of an effective lagrangian, we examine the effect of new
physics associated with third-generation fermions on $b \bar b$
production at the $Z$ resonance and study its implications for LEP2.  We
obtain the constraints on such operators with derivative couplings at
LEP1 and the SLC, and determine the prospects for detecting their
presence at LEP2. We find that despite the small observed deviations
from the standard model on the $Z$ resonance, much larger deviations
from standard model predictions are possible at LEP2, and the
corresponding couplings can be determined rather precisely.

\end{abstract}
              
\narrowtext

\vfill
\eject

\section{Introduction}

Although the Standard Model (SM) has been tremendously successful in
describing the physics of the electroweak interactions \cite{SM}, it is
quite possible that it is only an effective theory which at higher
energies will break down as the deeper structure of the underlying
physics emerges. There are reasons to believe that the deviation from
the standard model might first appear in the interactions involving the
third-generation fermions\cite{third}. The large value of the top quark
mass\cite{top}, on the order of the Fermi scale, means that the top
quark couples significantly to the electroweak symmetry-breaking sector.
Considerations of ``triviality'' \cite{triv} and ``naturalness''
\cite{nat} suggest that the scalar sector of the full theory may be more
complicated than the one in the standard model. If so, the Higgs sector of the
SM may be described by an effective theory which will most likely induce
new physics interactions for the third-generation fermions. In fact,
there are subtle deviations from standard model predictions for
measurements involving the $b$-quark on the $Z$ resonance \cite{data},
which, if they persist, could be an indication of new physics beyond the
standard model.

In this paper we examine the new physics effects on $b \bar b$
production at the $Z$ resonance and its implications for LEP2 in terms
of an effective lagrangian.  Specifically, we focus on a class of
CP-conserving dimension-six operators which can contribute to $e^+e^-
\rightarrow b \bar b$, and which involve a derivative coupling. Such
operators in principle can be detected at LEP1 and the SLC, and their
influence will increase at higher energies. We determine the limits on
the operators which contribute to $e^+e^- \rightarrow b \bar b$ from the
precision data at LEP1 and the SLC, and from partial-wave unitarity
considerations. We find that the data slightly favors nonzero values of
the anomalous couplings. Then we determine the range of possible
measurements at LEP2 allowed by these constraints, and show how one
might determine which operators are present if deviations from SM
predictions are observed at LEP2.

There are two kinds of effective lagrangians used in the literature to
describe the new physics of the top quark. In the so-called linear
realization, an explicit Higgs doublet of the SM is included and
electroweak symmetry breakdown is assumed to be induced by the Higgs
potential as in the SM. In this effective lagrangian, the leading terms
are given by the SM and the corrections which come from the underlying
theory beyond the SM are described by higher dimension operators.  So
the SM will be recovered in the limit $\Lambda \rightarrow \infty$ and
the mass of the Higgs boson is constrained by the "vacuum stability"
\cite{datta} and "triviality" \cite{chivu} arguments.  In the second
common approach, the effective lagrangian with a non-linearly
realization of the SM symmetry, one constructs non-linear fields from
the fermions, such as the top quark, and the would-be goldstone bosons
which give masses to the W and Z bosons. Without an explicit Higgs
doublet in the model, electroweak symmetry breakdown in the nonlinear
realization is generally assumed, but no specific mechanism needs to be
given.  In this effective lagrangian the leading terms are again the SM,
and the corrections, which come from the underlying theory responsible
for the symmetry breaking and mass generation, are described by higher
dimension operators and also by dimension-4 operators, which is ${\it
different}$ from the linear realization of the effective lagrangian.  A
Higgs field, as a singlet scalar, can be easily added to the lagrangian,
but its couplings to the matter fields will not be constrained by the
mass of the latter.

In this paper we take an effective lagrangian with a linear realization
of the SM symmetry. In Refs.~\cite{GPR} and \cite{wyyz}, the higher
dimension operators which involve fermions of the third generation,
gauge bosons, and Higgs bosons and which conserve CP are given. Here
we will restrict the list of the higher dimension operators to those 
which lead to changes in $e^+e^- \rightarrow b \bar b$ (since these are
accessible on the $Z$ resonance) and involve a derivative coupling.
These operators are
\begin{eqnarray}
\Delta {\cal L} = {1\over\Lambda^2} [
&c_{1L} g_2 \bar\Psi_L  \gamma_\mu \tau_a \Psi_L (D_\nu W^{\mu\nu})^a
+ c_{3L} g_1 \bar\Psi_L  \gamma_\mu \Psi_L D_\nu B^{\mu\nu}
+ \tilde c_{3R} g_1 \bar b_R  \gamma_\mu b_R D_\nu B^{\mu\nu}
\nonumber\\
&+ (c_{bW} g_2 \bar\Psi_L\sigma^{\mu\nu} \tau_a \Phi b_R W^a_{\mu\nu}
+ c_{bB} g_1 \bar\Psi_L\sigma^{\mu\nu} \Phi b_R B_{\mu\nu}
+ c_{Db} \bar\Psi_L D_\mu b_R D^\mu \Phi + H.c.)].
\label{eq:Lag}
\end{eqnarray}
where $\Psi_L = ( t, b)_L$ and $c_{1L}, c_{3L},$~etc., are free parameters
determining the strength of the corresponding operators in the effective
lagrangian. The contributions of these operators will be enhanced
relative to the standard model in processes with higher CM energy than
LEP1 and the SLC, such as at LEP2.

This paper is organized as follows. In Sec. II we will examine the
current constraints on these operators from LEP1 and the SLC, and in
Sec. III we will find the limits on these operators from partial wave
unitarity. In Sec. IV we will determine their possible effects at LEP2.
A brief discussion is given in Sec. V.  Finally, details for obtaining
the bounds from partial wave unitarity are presented in the Appendix.

\section{Constraints on the $Z$ resonance}

The effective Lagrangian for $e^+e^- \rightarrow b \bar{b}$ which arises
from the Standard Model plus the anomalous interactions listed above can
be written
\begin{eqnarray}
{\cal L}^{eff} = &\phantom{+}
\bar e_L \gamma_\mu e_L
\bar{b}\left[ G_{LL}\gamma^\mu \Gamma_L + G_{LR}\gamma^\mu \Gamma_R
+ G_{LS}{p_1^\mu - p_2^\mu\over\sqrt{s}}
- i G_{LT}\sigma^{\mu\nu}{p_1^\nu + p_2^\nu\over\sqrt{s}} \right] b
\nonumber\\
&+\bar e_R \gamma_\mu e_R
\bar{b}\left[ G_{RL}\gamma^\mu \Gamma_L + G_{RR}\gamma^\mu \Gamma_R
+ G_{RS}{p_1^\mu - p_2^\mu\over\sqrt{s}}
- i G_{RT}\sigma^{\mu\nu}{p_1^\nu + p_2^\nu\over\sqrt{s}} \right] b,
\end{eqnarray}
where (for convenience, a
momentum instead of a partial derivative in the effective lagrangian
is used) $p_1$ and $p_2$ refer to the momenta of the final state $b$
and $\bar b$ quarks, respectively. We note that although the terms
involving $\bar b (p_1^\mu - p_2^\mu) b$ can be transformed into
a linear combination of the other terms in ${\cal L}^{eff}$ using
the Gordon decomposition, we keep them here since they arise from
independent anomalous operators with different top quark couplings.
Also, terms involving $\bar b (p_1^\mu + p_2^\mu) b$ and $\bar b
(-i\sigma^{\mu\nu})(p_1^\nu - p_2^\nu) b$ (which are equal to each
other by the Gordon decomposition) are not included since their
contributions to the matrix element are proportional to the electron
mass and therefore negligible.

In terms of the couplings defined above
\begin{eqnarray}
G_{LL} =&
{e^2\over s} \left[ {1\over3} + {s\over\Lambda^2}(-c_{1L}+c_{3L}) \right]
+ {g_Z^2(-{1\over2}+s_W^2)\over s-m_Z^2+im_Z\Gamma_Z} \left[
-{1\over2}+{1\over3}s_W^2+{s\over\Lambda^2}(c_W^2 c_{1L}+s_W^2c_{3L})\right]
\label{eq:GLL}\\
G_{LR} =&
{e^2\over s} \left[ {1\over3} + {s\over\Lambda^2} \tilde c_{3R} \right]
+ {g_Z^2(-{1\over2}+s_W^2)\over s-m_Z^2+im_Z\Gamma_Z} \left[
{1\over3}s_W^2+{s\over\Lambda^2} s_W^2 \tilde c_{3R}\right]
\label{eq:GLR}\\
G_{RL} =&
{e^2\over s} \left[ {1\over3} + {s\over\Lambda^2}(-c_{1L}+c_{3L}) \right]
+ {g_Z^2(s_W^2)\over s-m_Z^2+im_Z\Gamma_Z} \left[
-{1\over2}+{1\over3}s_W^2+{s\over\Lambda^2}(c_W^2 c_{1L}+s_W^2c_{3L})\right]
\label{eq:GRL}\\
G_{RR} =&
{e^2\over s} \left[ {1\over3} + {s\over\Lambda^2} \tilde c_{3R} \right]
+ {g_Z^2(s_W^2)\over s-m_Z^2+im_Z\Gamma_Z} \left[
{1\over3}s_W^2+{s\over\Lambda^2} s_W^2 \tilde c_{3R}\right]
\label{eq:GRR}\\
G_{LS} =& {\sqrt{s} v_o\over \sqrt2\Lambda^2}
{g_Z^2(-{1\over2}+s_W^2)\over s-m_Z^2+im_Z\Gamma_Z} [c_{Db}]
\label{eq:GLS}\\
G_{RS} =& {\sqrt{s} v_o\over \sqrt2\Lambda^2}
{g_Z^2(s_W^2)\over s-m_Z^2+im_Z\Gamma_Z} [c_{Db}]
\label{eq:GRS}\\
G_{LT} =& {\sqrt{2s} v_o\over \Lambda^2} \left\{
{e^2\over s} \left[ -c_{bW}+c_{bB}) \right]
+ {g_Z^2(-{1\over2}+s_W^2)\over s-m_Z^2+im_Z\Gamma_Z} \left[
c_W^2 c_{bW}+s_W^2c_{bB}\right]
\right\}
\label{eq:GLT}\\
G_{RT} =& {\sqrt{2s} v_o\over \Lambda^2} \left\{
{e^2\over s} \left[ -c_{bW}+c_{bB}) \right]
+ {g_Z^2(s_W^2)\over s-m_Z^2+im_Z\Gamma_Z} \left[
c_W^2 c_{bW}+s_W^2c_{bB}\right]
\right\},
\label{eq:GRT}\end{eqnarray}
where $s_W = \sin\theta_W$, $c_W = \cos\theta_W$, and $g_Z = e/(s_W c_W)$.
Then for the process $e^+e^- \rightarrow b \bar b$ we find
\begin{eqnarray}
\sigma^b_F + \sigma^b_B =& {s\over16\pi} \left\{
|G_{LL}|^2 + |G_{LR}|^2 + |G_{RL}|^2 + |G_{RR}|^2 + |G_{LS}+G_{LT}|^2
+ |G_{RS}+G_{RT}|^2 \right.
\nonumber\\
&\left. - {2m_b\over\sqrt{s}}{\cal R}e \left[
(G_{LL}+G_{LR})(G_{LS}+3G_{LT})^* + (G_{RL}+G_{RR})(G_{RS}+3G_{RT})^*
\right]\right\},
\label{eq:F+B}
\\
\sigma^b_F - \sigma^b_B =& {3s\over64\pi} \left\{
|G_{LL}|^2 - |G_{LR}|^2 - |G_{RL}|^2 + |G_{RR}|^2 \right.
\nonumber\\
&\left. -{4m_b\over\sqrt{s}}{\cal R}e \left[
(G_{LL}-G_{LR})G_{LT}^* - (G_{RL}-G_{RR})G_{RT}^*
\right]\right\},
\label{eq:F-B}
\end{eqnarray}
where $\sigma^b_F$ and $\sigma^b_B$ are the forward and backward
cross sections, and we have kept terms only up to first order in
$m_b/\sqrt{s}$. We note that in Eqs.~\ref{eq:F+B} and \ref{eq:F-B}
there are $1/\Lambda^4$ terms, and that for completeness one should in
principle also include dimension-eight operators which can contribute to
the process and are the same
order in $\Lambda$. We do not include them here as we only wish to
illustrate the importance and feasibility of studying such anomalous
couplings.

On the $Z$ resonance only the $Z$ couplings contribute appreciably and
we can define
\begin{equation}
G_{ij} \approx {g_Z^2\over im_Z\Gamma_Z} g^e_i g^b_j, \qquad i,j=L,R,S,T,
\end{equation}
where
\begin{eqnarray}
g^e_L = -{1\over2} + s_W^2 \ ,& \ g^e_R = s_W^2,
\label{eq:gLRe}
\\
g^b_L = -{1\over2} + {1\over3}s_W^2 
+ {m_Z^2\over\Lambda^2} (c_W^2 c_{1L} + s_W^2 c_{3L})
\ ,& \
g^b_R = {1\over3} s_W^2
+ {m_Z^2\over\Lambda^2} s_W^2 \tilde c_{3R},
\label{eq:gLRb}
\\
g^b_S = {m_Z v_o\over\sqrt2\Lambda^2} c_{Db}
\ ,& \
g^b_T = {\sqrt{2}m_Zv_o\over\Lambda^2} (c_W^2 c_{bW} + s_W^2 c_{bB}).
\label{eq:gSTb}
\end{eqnarray}

The measurables involving the $b$ couplings on the $Z$ resonance are the
partial decay width
\begin{equation}
{\Gamma_b\over\Gamma_b^{SM}} =
{\left[ (g_L^b)^2 + (g_R^b)^2 + (g_T^b + g_S^b)^2
- {2m_b\over m_Z}(g_L^b + g_R^b)(3g_T^b + g_S^b) \right]
\over
[(g_L^b)^2+(g_R^b)^2]_{SM}},
\end{equation}
and the $b$-quark asymmetry
\begin{equation}
A_b =
{\left[ (g_L^b)^2 - (g_R^b)^2 - {4m_b\over m_Z}(g_L^b-g_R^b)g_T^b
\right]
\over
\left[ (g_L^b)^2 + (g_R^b)^2 + (g_T^b + g_S^b)^2
- {2m_b\over m_Z}(g_L^b + g_R^b)(3g_T^b + g_S^b) \right]},
\end{equation}
where again we have kept terms up to first order in $m_b/\sqrt{s}$. The
quantity $\Gamma_b/\Gamma_b^{SM}$ can be determined from $R_b =
\Gamma_b/\Gamma_{had} = (\Gamma_b/\Gamma_b^{SM})/[(\Gamma_b/\Gamma_b^{SM})
+(1/R_b^{SM}) - 1]$, assuming that the only deviation of the $Z$ width
from the SM occurs in $\Gamma_b$, and $A_b$ can be determined from
$A_{FB}^b = {3\over4}A_eA_b$ and
$A_e = [(g_L^e)^2 - (g_R^e)^2]/[(g_L^e)^2 + (g_R^e)^2]$. The latest
experimental measurements\cite{data} are
\begin{equation}
R_b = 0.2178 \pm 0.0011 \ , \ A_b = 0.883 \pm 0.025,
\label{eq:data}
\end{equation}
where the Standard Model values are
\begin{equation}
R_b^{SM} = 0.2156 \ , \ A_b^{SM} = 0.936,
\label{eq:sm}
\end{equation}
when $\sin^2\theta_W=0.232$. Both the $R_b$ and $A_b$ measurements are
about 2-sigma from the standard model. The value for $A_b$ in
Eq.~\ref{eq:data} is determined by combining the LEP and SLC results in
quadrature; the LEP value is determined from the LEP measurements of
$A_{FB}^b$ and $A_e$, while the SLC value is measured directly using a
forward-backward polarization asymmetry. A different LEP value of $A_b$
can be found by using the LEP value for $A_{FB}^b$ and the world average
$A_e$, which then gives\cite{data}
\begin{equation}
A_b = 0.867\pm0.022,
\label{eq:data2}
\end{equation}
for the combined measurement. This alternative value of $A_b$ is about
3-sigma away from the standard model value given in Eq.~\ref{eq:sm}.
We will not use the value in Eq.~\ref{eq:data2} in our analysis below,
but will discuss briefly its consequences in Sec.~V.



In Figs.~1 and 2 we show the regions allowed by the LEP1 and SLC data in
Eq.~\ref{eq:data} at 95\%~CL for various subsets
of the parameters in the effective Lagrangian in Eq.~\ref{eq:Lag}.
Six of the ten parameters affect $e^+e^- \rightarrow b\bar b$. Fig.~1
shows the allowed region for the parameters $c_{3L}$ and the linear
combination $\tilde c_{3R}+c_{1L}c_W^2/s_W^2$, and Fig.~2 for $c_{Db}$
and the linear combination $c_{bB}+c_{bW}c_W^2/s_W^2$. The four
parameters $c_{3L}$, $\tilde c_{3R}$, $c_{Db}$, and $c_{bB}$ are
suppressed by a factor $s_W^2$ on the $Z$ resonance,
but not in their couplings to the photon. Hence they can give
large effects at higher energies, such as at LEP2, where the photon
contribution becomes important. Furthermore, the leading contributions
of all six parameters which contribute to $e^+e^- \rightarrow b \bar b$
will be enhanced by a factor of $s/m_Z^2$ at higher energies compared
to the $Z$ resonance.




\section{Limits from partial wave unitarity}

The unitarity limits on anomalous third-family couplings not constrained
by $e^+e^- \rightarrow b\bar b$ have been calculated in Ref.~\cite{GPR}.
Here we determine the unitarity limits on all six anomalous couplings in
Eq.~\ref{eq:Lag}; the details of the calculation are given in the
Appendix. Despite the strong constraints from the data on the
$Z$ resonance, the energy dependence of these couplings implies that
there will be significant constraints from unitarity as well. For
$c_{1L}$, $c_{3L}$ and $\tilde c_{3R}$, the best limit from unitarity
comes from the $J=1$ partial wave amplitude for the 2-to-2 processes
involving the helicity channels $b_+ \bar b_-$, $b_- \bar b_+$,
$t_+ \bar t_-$, and $t_- \bar t_+$. The largest eigenvalue of the
$4\times4$ coupled-channel matrix leads to the strongest constraint.
Although the exact expressions are somewhat messy, if we keep only the
terms quadratic in the anomalous couplings that are enhanced by a factor
$s^2/\Lambda^4$ at high energies, the unitarity condition can be written
approximately as
\begin{eqnarray}
\tilde c_{3R}^2 + 2c_{3L}^2 < {2c_W^2\over\alpha} {\Lambda^4\over s^2}
\approx \left({14~TeV^2\over s}\right)^2,
\label{eq:unit1} \\
|c_{1L}| < {s_W\over\sqrt\alpha} {\Lambda^2\over s}
\approx {5.5~TeV^2\over s},
\label{eq:unit2}
\end{eqnarray}
where we have used $\Lambda=1$~TeV. For $c_{Db}$, $c_{bB}$ and $c_{bW}$,
the best limit from unitarity comes from the $J=0$ partial wave
amplitude for the processes involving the channels $b_+ \bar b_+$ and
$b_- \bar b_-$. The largest eigenvalue of the $2\times2$ coupled-channel
matrix leads to the strongest constraint. Keeping only the leading terms
for each coupling, we find the constraint
\begin{equation}
34 c_{Db}^2 +
9 {m_Z^2\over s} [8 (c_{bB} s_W^2 + c_{bW} c_W^2) - c_{Db}]^2
< {768 s_W^2 c_W^2\over \alpha} {m_Z^2\over v_o^2} {\Lambda^4\over s^2}
\approx \left({49 TeV^2\over s}\right)^2.
\label{eq:unit3}
\end{equation}
The unitarity constraints for $\Lambda=1$~TeV are shown in Figs.~3 and 4 for
$\sqrt{s}=1$, $2$, and $3$~TeV. The limits from LEP1 and the SLC are
also shown for comparison.

\section{Effects on $b \bar b$ Production at LEP2}

The next step is to determine the range of measurements at LEP2 that are
possible which are also consistent with the LEP and SLC data. From
Eqs.~\ref{eq:GLL} to \ref{eq:F-B} we note that to leading order in $s$,
the anomalous interactions have contributions proportional to
$s/\Lambda^2$ (from the interference of SM and anomalous pieces in
$G_{LL}$, $G_{LR}$, $G_{RL}$, and $G_{RR}$) and $sv_o^2/\Lambda^4$ (from
$G_{LS}$, $G_{LS}$, $G_{LT}$, and $G_{RT}$).  These contributions can
therefore lead to large deviations from the SM at higher energies. In
our analysis, we study the operators which involve vector and axial
vector couplings ($c_{1L}$, $c_{3L}$, and $\tilde c_{3R}$) separately
from those which introduce scalar and tensor couplings ($c_bB$,
$c_{bW}$, and $c_{Db}$). In Fig.~5 we show the range of measurements
possible for $\sigma_b \equiv \sigma(e^+e^-\rightarrow b\bar b)$ and
$A_{FB,b} \equiv (\sigma_{F,b} - \sigma_{B,b})/(\sigma_{F,b} +
\sigma_{B,b})$ at LEP2, given the constraints on the parameters from LEP
and SLC data and partial wave unitarity with the new physics scale set
at 1~TeV.

Fig.~5a shows the possible range of measurements at LEP2 with $\sqrt{s}=
170$~GeV for $\Delta A_{FB}/A_{FB} \equiv
(A_{FB,b}-A_{FB,b}^{SM})/A_{FB,b}^{SM}$ and $\Delta\sigma/\sigma \equiv
(\sigma_b-\sigma_b^{SM})/\sigma_b^{SM}$ when $c_{3L}$ and $\tilde
c_{3R}$ are varied over their allowed values, for $c_{1L}=1$, $0$, and
$-1$ (larger values of $c_{1L}$ are not allowed by unitarity). The
horizontal grid lines correspond to constant values of $c_{3L}$, while
the vertical grid lines correspond to constant values of $\tilde
c_{3R}$. We see that sizeable deviations from the standard model are
possible, up to 60\% in $\sigma$ and 40\% in $A_{FB}$. Even larger
deviations are possible if $A_b$ from Eq.~\ref{eq:data2} is used. The
corresponding ranges for $\sqrt{s}=190$ and $200$~GeV are shown in
Figs.~5b and 5c, respectively. A comparison of the graphs in Fig.~5
shows that the deviations grow with $\sqrt{s}$, as expected from the $s$
dependence of the anomalous contributions.

In principle one should be able to deduce the three values of the
anomalous parameters $c_{1L}$, $c_{3L}$, and $\tilde c_{3R}$ from the
combined measurements of $\sigma_b$ and $A_{FB}^b$ taken at different
energies. As an illustration, let's assume that the measurements
$\Delta\sigma/\sigma= 0.07$, $0.09$, $0.11$, and $\Delta A_{FB}/A_{FB}=
-0.08$, $-0.10$, $-0.11$ are taken at $\sqrt{s}= 170$, $190$, and
$200$~GeV, respectively. Assuming an integrated luminosity of 10
fb$^{-1}$ at each energy, the standard statistical errors are
$\delta\sigma_b/\sigma_b = 1/\sqrt{N} \approx .04$, and
$\delta A_{FB,b} = 2\sqrt{N_F N_B/N^3} \approx .04$, where $N_F$ and
$N_B$ are the number of forward and backward $b\bar b$ events,
respectively, and $N=N_F+N_B$. Then the best least-squares fit to
the anomalous parameters is $c_{1L}=0.0\pm0.2$,
$c_{3L}=-1.70^{+0.28}_{-0.13}$, and $\tilde c_{3R}= 2.70^{+0.16}_{-0.65}$.
From this exercise it is clear that the
combined measurements of $\sigma_b$ and $A_{FB}^b$ taken at different
energies at LEP2 is sufficient to greatly constrain the parameter space,
and, if the deviations from the standard model are large enough, provide
strong evidence for new physics involving the third generation fermions.

We have done a similar analysis for the parameters involving the scalar
and tensor interactions, $c_{bB}$, $c_{bW}$, and $c_{Db}$. In Fig.~6a we
show the possible range of measurements at LEP2 with $\sqrt{s}= 170$~GeV
for $\Delta A_{FB}/A_{FB}$ and $\Delta\sigma/\sigma$ when $c_{bB}$ and
$c_{Db}$ are varied over their allowed values, for $c_{bW}=2$,
$0$, and $-2$ (larger values of $c_{bW}$ are not allowed by
unitarity). Fig.~6b shows the corresponding ranges for $c_{bW}=1$ and
$-1$. We see from Fig.~6 that the range of deviations from the
standard model is not as large as in Fig.~5, and there is more 
ambiguity as to which combinations of parameters yield a certain set of 
measurements. Also, for the parameters which introduce scalar and tensor
couplings there is a weaker $\sqrt{s}$ dependence, which makes a precise
determination of their exact values more difficult than the case
involving vector and axial vector couplings.

\section{Discussion}
%
%

We have found the constraints on a class of six dimension-6 operators
involving $b$-quarks with derivative couplings from LEP and SLC data and
from partial wave unitarity. We have also shown that these operators may
lead to large deviations from the standard model predictions for the
total cross section and forward-backward asymmetry in $b\bar b$
production at LEP2. The effect of the dimension-six operators with
derivative coupling at LEP2 has been discussed in Ref.~\cite{wyyz}, and
the three vector and axial vector operators have also been studied in
detail in Ref.~\cite{Gounaris2} for LEP2 and NLC energies. In our
analysis we show that precise measurements of the $b \bar b$ cross
section together with the forward-backward asymmetry {\it at different
energies} can greatly limit the possible range of anomalous couplings
which might be responsible for the deviations. If these anomalous
operators are in fact present, even larger deviations can be expected
at the much higher energy scale of the NLC \cite{Gounaris2}.

\section{Acknowledgements}

We thank A. Firestone and E. Rosenberg for discussions.
This work was supported in part by the U.S.~Department of Energy under
Contracts DE-FG02-94ER40817 and DE-FG02-92ER40730. XZ was supported in
part by the National Sceince foundation of China.

\section{Appendix: Bounds from partial wave unitarity}

In this Appendix we describe in more detail the derivation of the bounds
on the anomalous couplings from partial wave unitarity. In each case we
consider contributions which are quadratic in the anomalous couplings,
and treat the couplings for the vector and axial vector interactions
separately from those for the scalar and tensor interactions.

For the parameters $c_{1L}$, $c_{3L}$, and $\tilde c_{3R}$, the best
limit from partial wave unitarity comes from the color singlet $J=1$
amplitudes. The $J=1$ helicity amplitudes for the processes which are
quadratic in the anomalous couplings are, to leading order in $s$,
\begin{eqnarray}
T_1(b_+ \bar b_- \rightarrow b_+ \bar b_-)=&
{11\over24} {\alpha\over c_W^2} {s^2\over\Lambda^4} \tilde c_{3R}^2,
\\
T_1(t_+ \bar t_- \rightarrow t_+ \bar t_-)=& 
T_1(b_+ \bar b_- \leftrightarrow t_+ \bar t_-)= 0
\\
T_1(b_- \bar b_+ \rightarrow b_- \bar b_+)=&
T_1(t_- \bar t_+ \rightarrow t_- \bar t_+)=
{11\over24} {\alpha\over c_W^2} {s^2\over\Lambda^4}
\left( c_{3L}^2 + c_{1L}^2 {c_W^2\over s_W^2} \right),
\\
T_1(b_- \bar b_+ \leftrightarrow t_- \bar t_+)=&
{\alpha\over c_W^2} {s^2\over\Lambda^4}
\left( {1\over2} c_{3L}^2 - {7\over12}c_{1L}^2 {c_W^2\over s_W^2} \right),
\\
T_1(b_+ \bar b_- \leftrightarrow b_- \bar b_+)=&
T_1(b_+ \bar b_- \leftrightarrow t_- \bar t_+)=
- {1\over2} {\alpha\over c_W^2} {s^2\over\Lambda^4} c_{3L} \tilde c_{3R},
\\
T_1(t_+ \bar t_- \leftrightarrow b_- \bar b_+)=&
{1\over8} {\alpha\over c_W^2} {s^2\over\Lambda^4}
{m_t^2\over m_Z^2} c_{1L}^2 {c_W^2\over s_W^2},
\\
T_1(t_+ \bar t_- \leftrightarrow t_- \bar t_+)=&
{1\over16} {\alpha\over c_W^2} {s^2\over\Lambda^4}
{m_t^2\over m_Z^2} \left( c_{3L}^2 + c_{1L}^2 {c_W^2\over s_W^2} \right),
\\
\end{eqnarray}
where terms involving $m_b/m_Z$ or smaller have been dropped. The best
constraint comes from the eigenvalues of the coupled channel matrix. An
approximate solution can be found by ignoring the $m_t^2/m_W^2$ terms
(which are suppressed somewhat by small coefficients), and replacing the
terms with coefficients 11/24 and 7/12 by terms with coefficients of
1/2. Then the coupled channel matrix for the $J=1$ partial wave
amplitude in the $b_+ \bar b_-$, $t_+ \bar t_-$, $b_- \bar b_+$,
$t_- \bar t_+$ basis is
\begin{equation}
a_1 = {1\over2} {\alpha\over c_W^2} {s^2\over\Lambda^4} \pmatrix{
\tilde c_{3R}^2 & 0 & - c_{3L} \tilde c_{3R} & - c_{3L} \tilde c_{3R}
\cr
0 & 0 & 0 & 0
\cr
- c_{3L} \tilde c_{3R} & 0
& c_{3L}^2 + c_{1L}^2 {c_W^2\over s_W^2}
& c_{3L}^2 - c_{1L}^2 {c_W^2\over s_W^2} 
\cr
- c_{3L} \tilde c_{3R} & 0
& c_{3L}^2 - c_{1L}^2 {c_W^2\over s_W^2}
& c_{3L}^2 + c_{1L}^2 {c_W^2\over s_W^2} 
\cr}.
\label{eq:eig}
\end{equation}
The eigenvalues of the matrix in Eq.~\ref{eq:eig} are
\begin{equation}
\lambda = 0,\quad 0,\quad
{\alpha\over c_W^2} {s^2\over\Lambda^4} c_{1L}^2 {c_W^2\over s_W^2},\quad
{1\over2} {\alpha\over c_W^2} {s^2\over\Lambda^4}
(2 c_{3L}^2 + \tilde c_{3R}^2).
\end{equation}
The approximate partial wave unitarity bounds in Eqs.~\ref{eq:unit1} and
\ref{eq:unit2} are determined by requiring that the coupled channel
eigenvalues be less than unity.

For the parameters $c_{bB}$, $c_{bW}$, and $c_{Db}$, the best limit
from partial wave unitarity comes from the color singlet $J=0$
amplitudes. The $J=0$ amplitudes for the processes which are quadratic
in the anomalous couplings are, to leading order in $s$,

\begin{eqnarray}
T_o(b_+ \bar b_+ \rightarrow b_+ \bar b_+)=&
T_o(b_- \bar b_- \rightarrow b_- \bar b_-)=
-{3\over128} {\alpha\over s_W^2 c_W^2} {s^2\over\Lambda^4}
{v_o^2\over m_Z^2} c_{Db}^2,
\\
T(b_+ \bar b_+ \leftrightarrow b_- \bar b_-)=&
-{1\over768} {\alpha\over s_W^2 c_W^2} {s^2\over\Lambda^4}
\left[ 9{m_Z^2\over s} (8c_{bB} s_W^2 + 8 c_{bW} c_W^2 - c_{Db})^2
+ 16 c_{Db}^2 \right].
\end{eqnarray}
These may be summarized by the $J=0$ coupled channel matrix in the
$b_+ \bar b_+$,  $b_- \bar b_-$ basis
\begin{equation}
a_0 = -{1\over768} {\alpha\over s_W^2 c_W^2} {s^2\over\Lambda^4}
\pmatrix{
18 c_{Db}^2
& 9 {m_Z^2\over s} c^2 +16 c_{Db}^2
\cr
9 {m_Z^2\over s} c^2 +16 c_{Db}^2
& 18 c_{Db}^2
\cr},
\end{equation}
where $c \equiv 8 c_{bB} s_W^2 + 8 c_{bW} c_W^2 - c_{Db}$.
There are diagrams quadratic in the couplings which contribute to
processes such as $t \bar t \leftrightarrow b \bar b$, but the leading
term in $s$ for these amplitudes are $J=1$, and give a looser
constraint. There are also diagrams which are quadratic in couplings
that do not affect LEP1 and SLC measurables but contribute to processes
such as $t \bar t \leftrightarrow t \bar t$. Since in this paper we are
primarily interested in only operators constrained by LEP1 and the SLC, we
do not include the effects of these other operators.  As before, the
partial wave unitarity bound in Eq.~\ref{eq:unit3} is determined by
requiring that the maximum coupled channel eigenvalue be less than unity
(the smallest eigenvalue in this case always gives a looser constraint).

\vfill
\eject


\vfill
\eject


\centerline{FIGURE CAPTIONS}


\bigskip
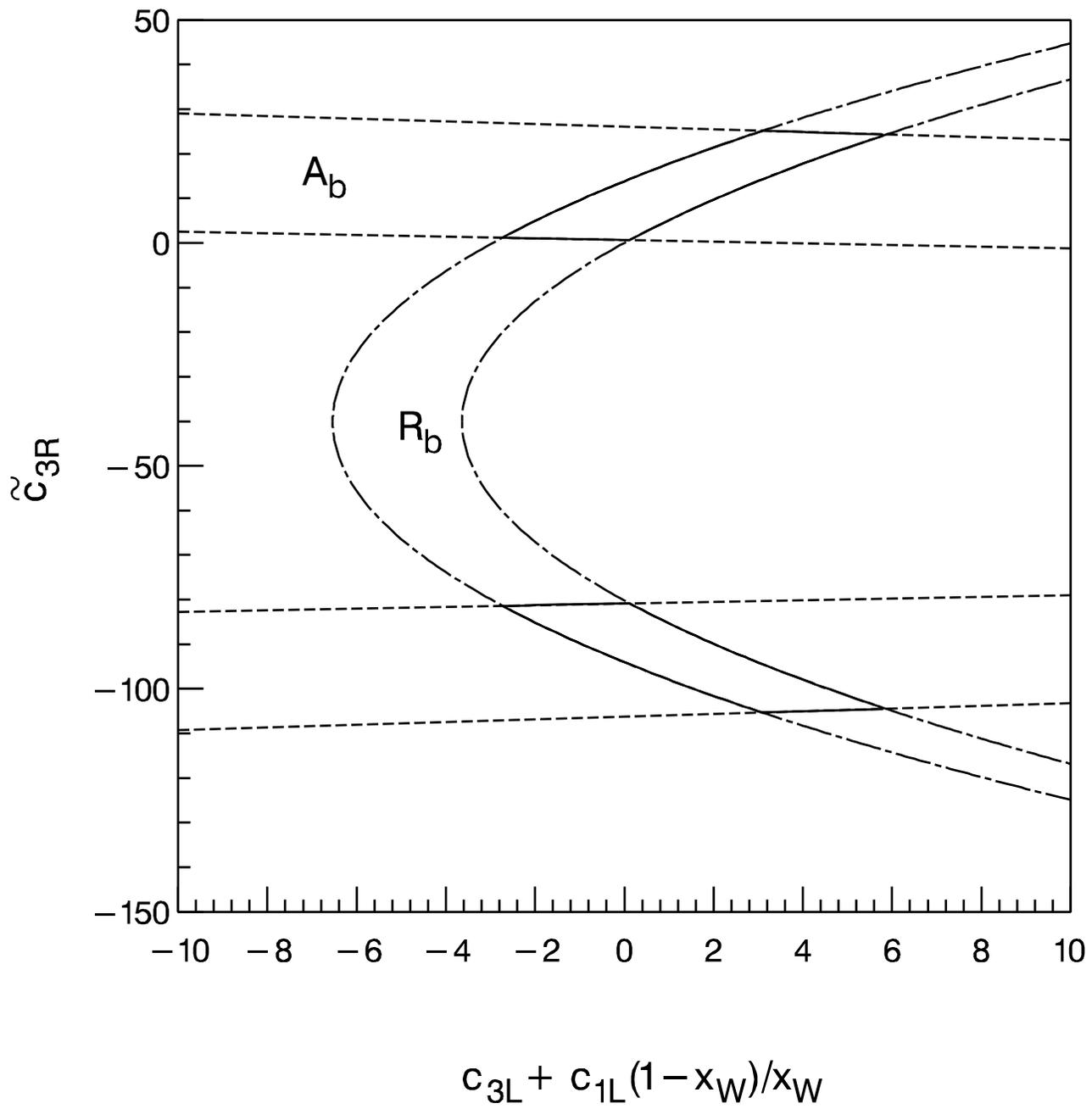
FIG.~1. 95\%~CL constraints from $R_b$ (dash-dotted lines) and $A_b$
(dashed lines) for the parameters $\tilde c_{3R}$ versus
$c_{3L}+c_{1L}c_W^2/s_W^2$. The regions where the bands overlap are
indicated by the solid lines. The data are taken from Ref.~\cite{data}.

\bigskip
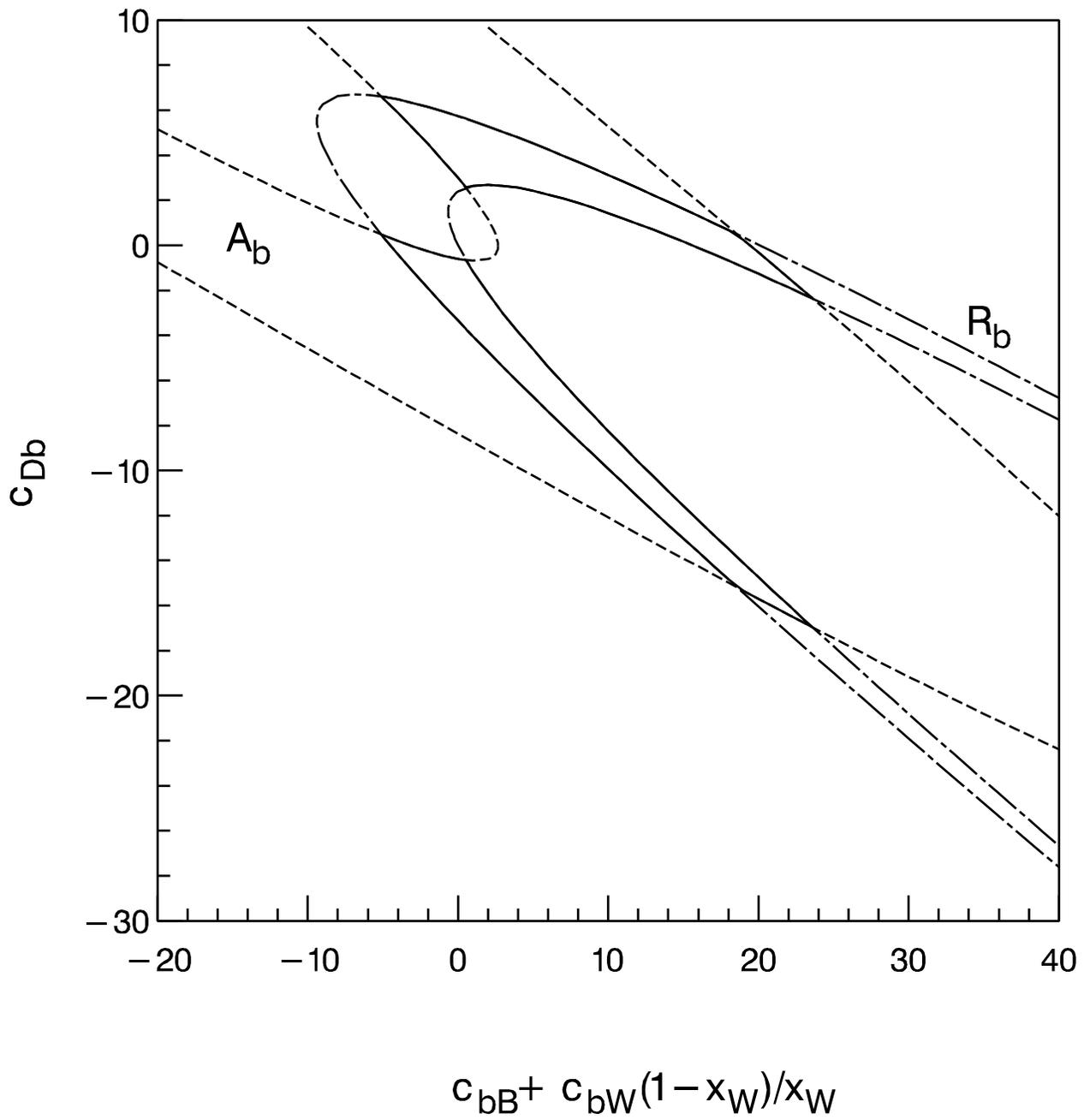
FIG.~2. 95\%~CL constraints from $R_b$ (dash-dotted lines) and $A_b$
(dashed lines) for the parameters $c_{Db}$ versus $c_{bB} + c_{bW}
c_W^2/s_W^2$. The notation and data used are the same as in Fig.~1.

\bigskip
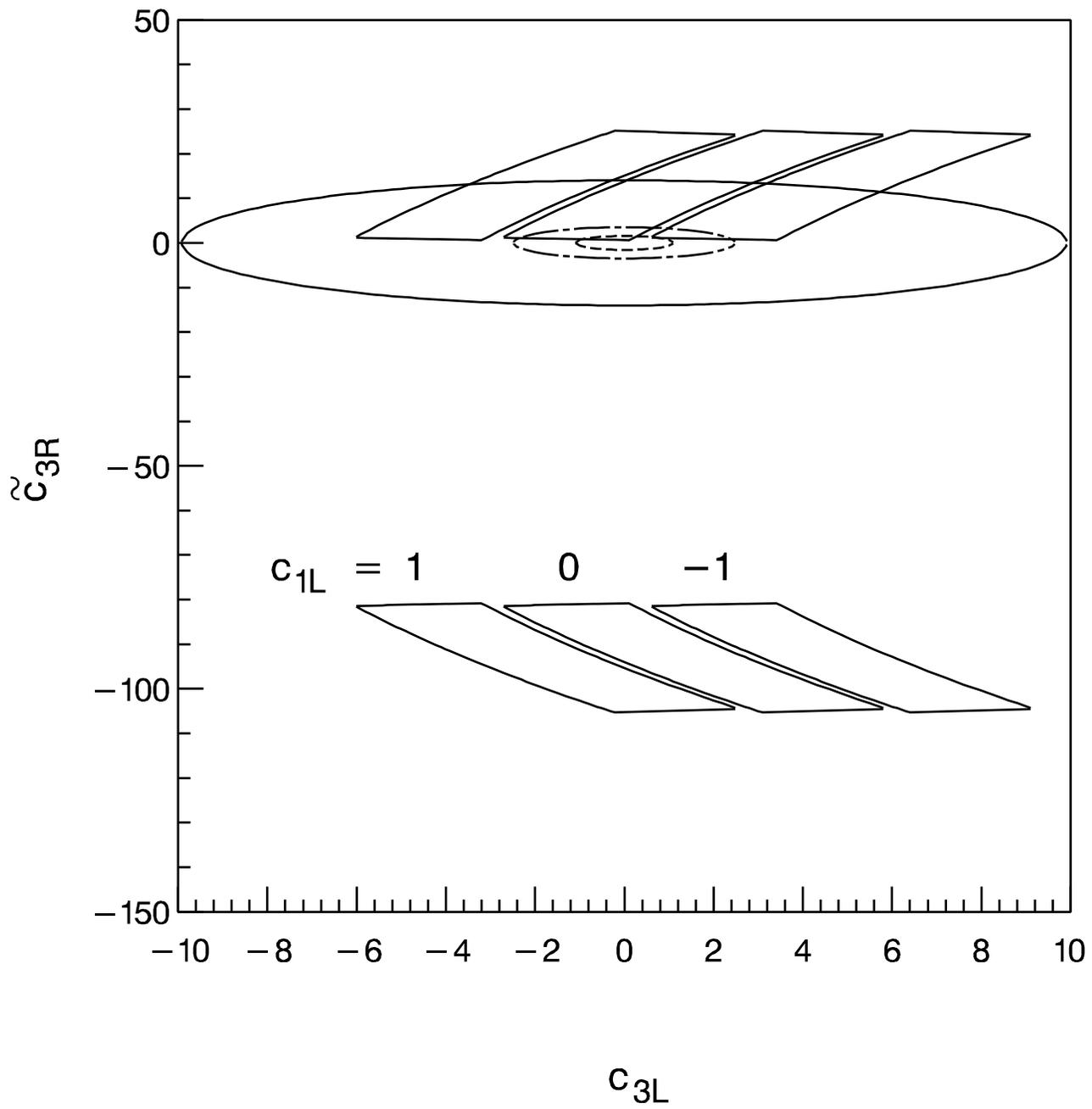
FIG.~3. Constraints from partial wave unitarity for $\sqrt{s}=$1~TeV
(solid lines), 2~TeV (dash-dotted lines), and 3~TeV (dashed lines) for
$\tilde c_{3R}$ versus $c_{3L}$, with $\Lambda=1$~TeV. The regions
inside the curves are allowed. The regions allowed by LEP and SLC data
at 95\%~CL, taken from Fig.~1, are also shown. The allowed regions are
shown for three different values of $c_{1L}$.

\bigskip
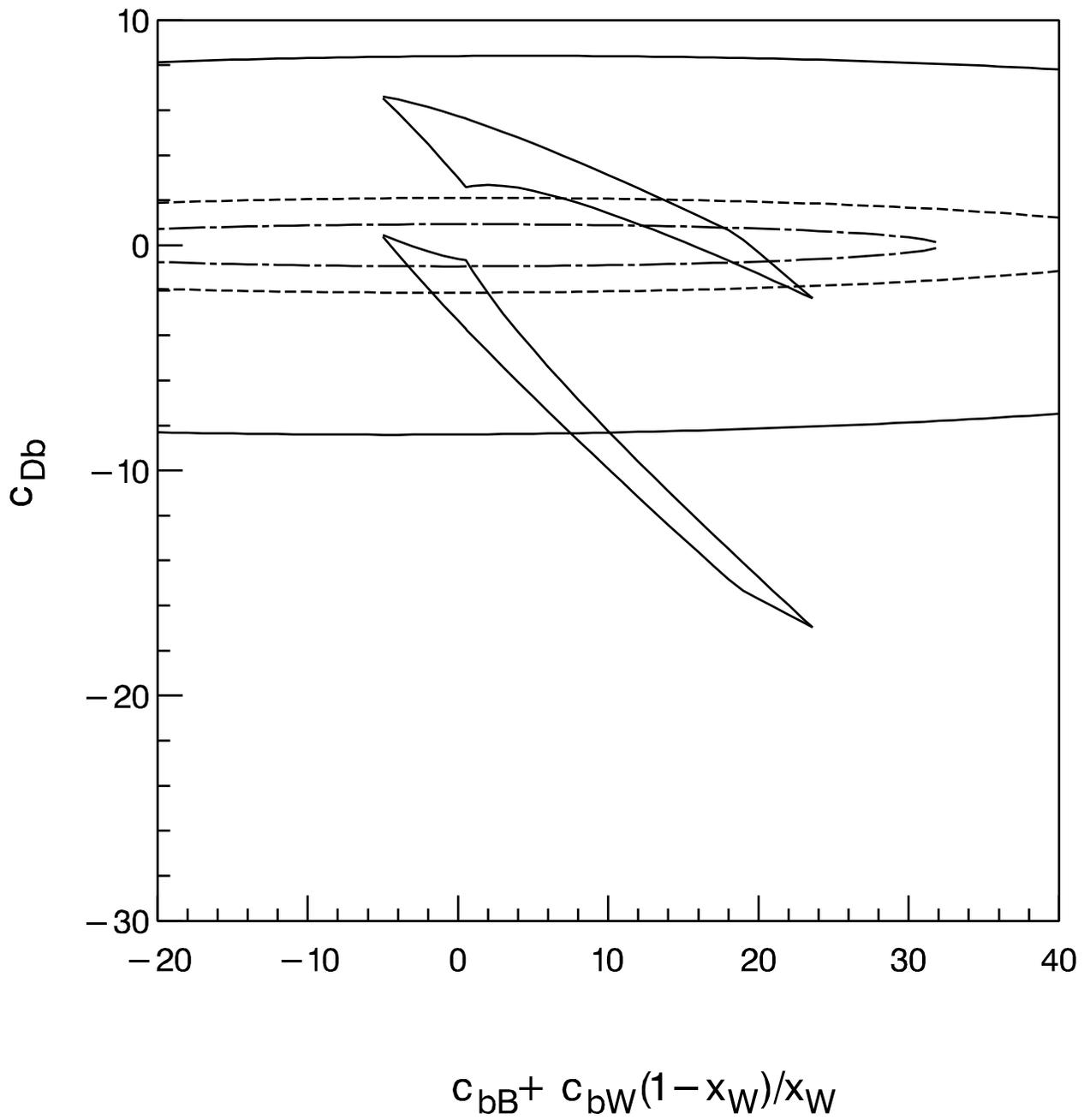
FIG.~4. Constraints from partial wave unitarity for $\sqrt{s}=$1~TeV
(solid lines), 2~TeV (dash-dotted lines), and 3~TeV (dashed lines) for
$c_{bB} + c_{bW}c_W^2/s_W^2$ versus $c_{Db}$, with $\Lambda=1$~TeV. The
regions inside the curves are allowed. The regions allowed by LEP and
SLC data at 95\%~CL, taken from Fig.~2, are also shown.

\bigskip
FIG.~5 Possible range of measurements for $\Delta A_{FB}/A_{FB}$ and
$\Delta\sigma/\sigma$ at LEP2 with (a) $\sqrt{s}=170$~GeV, (b)
$\sqrt{s}=190$~GeV, and (c) $\sqrt{s}=200$~GeV, given the constraints
from LEP, the SLC, and partial wave unitarity when $\sqrt{s}=\Lambda=1$~TeV.
The horizontal grid lines correspond to constant values of
$c_{3L}$, while the vertical grid lines correspond to constant
values of $\tilde c_{3R}$, while $c_{1L}$ is held fixed at $1$, $0$, and
$-1$. The points marked by a dot correspond to $c_{3L}=-1.7$ and
$\tilde c_{3R}=2.7$ in each case, and the space between grid lines
corresponds to a change of 0.1 and 0.5 in $c_{3L}$ and $\tilde c_{3R}$,
respectively.

\bigskip
FIG.~6 Possible range of measurements for $\Delta A_{FB}/A_{FB}$ and
$\Delta\sigma/\sigma$ at LEP2 with $\sqrt{s}=170$~GeV when $c_{bB}$ and
$c_{Db}$ are varied, given the constraints from LEP, the SLC, and partial
wave unitarity when $\sqrt{s}=\Lambda=1$~TeV. The ranges are shown for
(a) $c_{bW}=2.0$, $0.0$, and $-2.0$, and (b) $c_{bW}=1.0$ and $-1.0$.


\begin{references}
\bibitem{Gounaris2}
During preparation of this manuscript we received a preprint by
G.J. Gounaris, D.T. Papadamou, and F.M. Renard, preprint hep-ph/9703281,
March 1997, which also examines the effect of the operators moderated
by the couplings $c_{1L}$, $c_{3L}$ and $\tilde c_{3R}$.
\end{references}
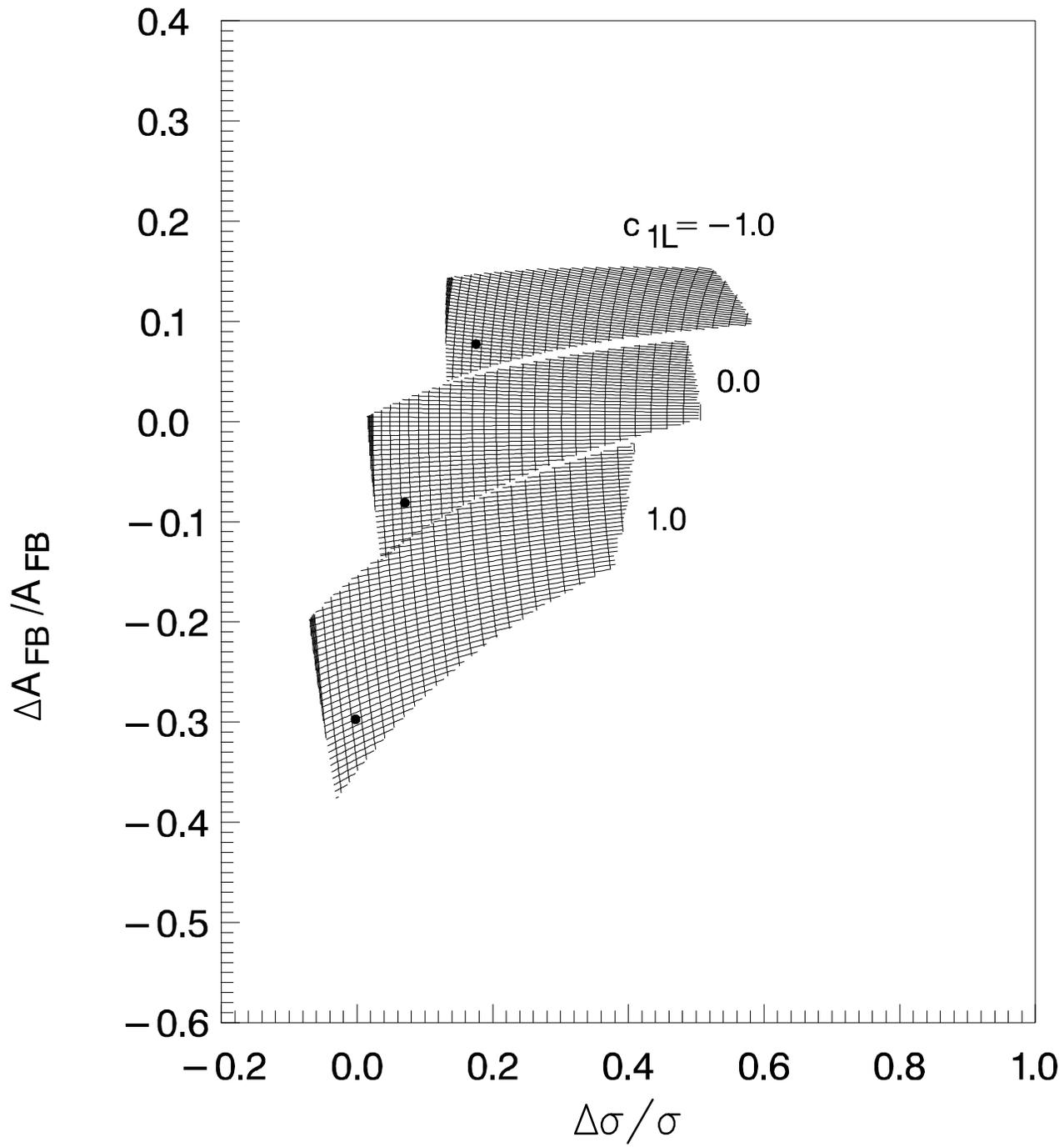
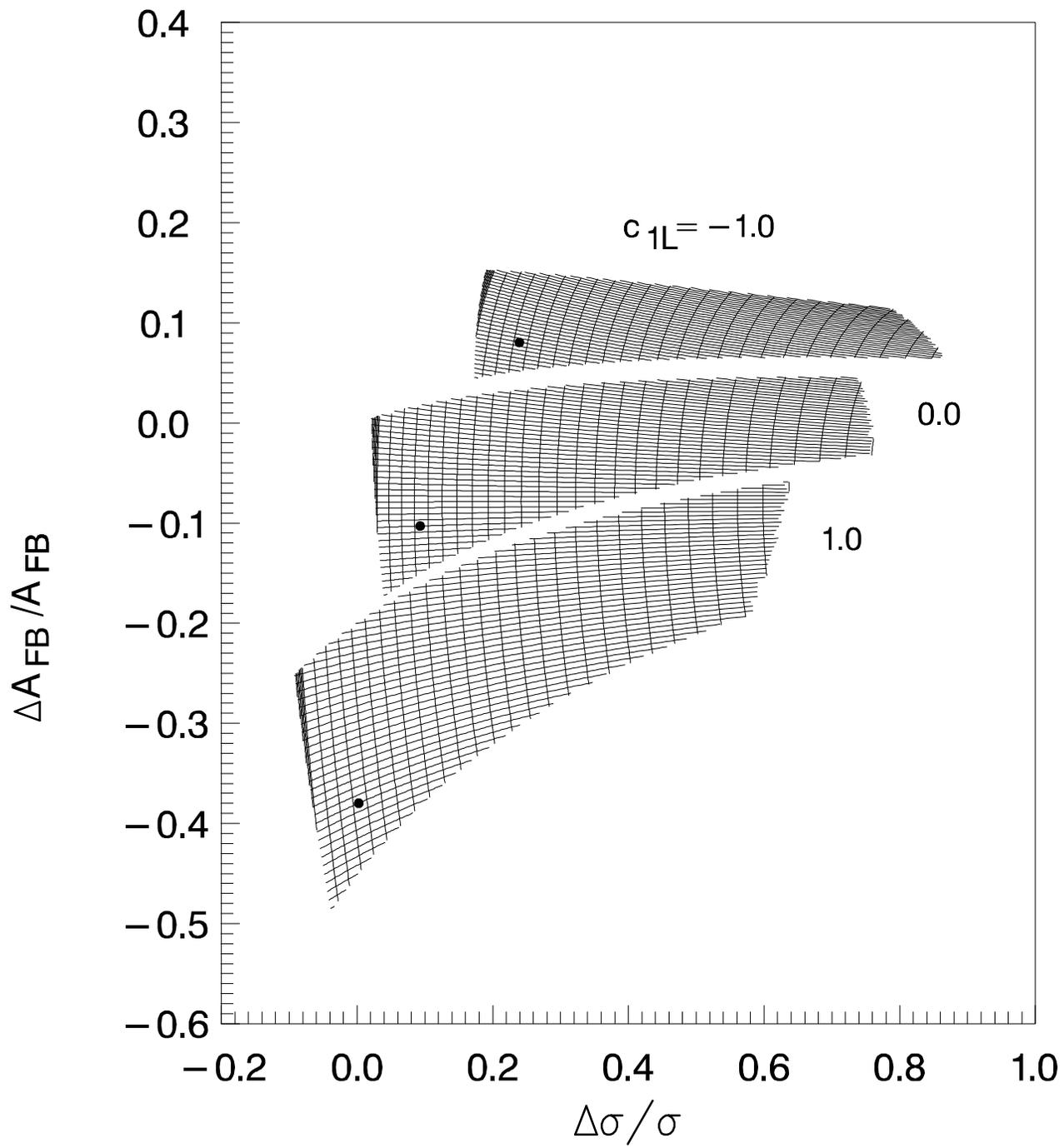
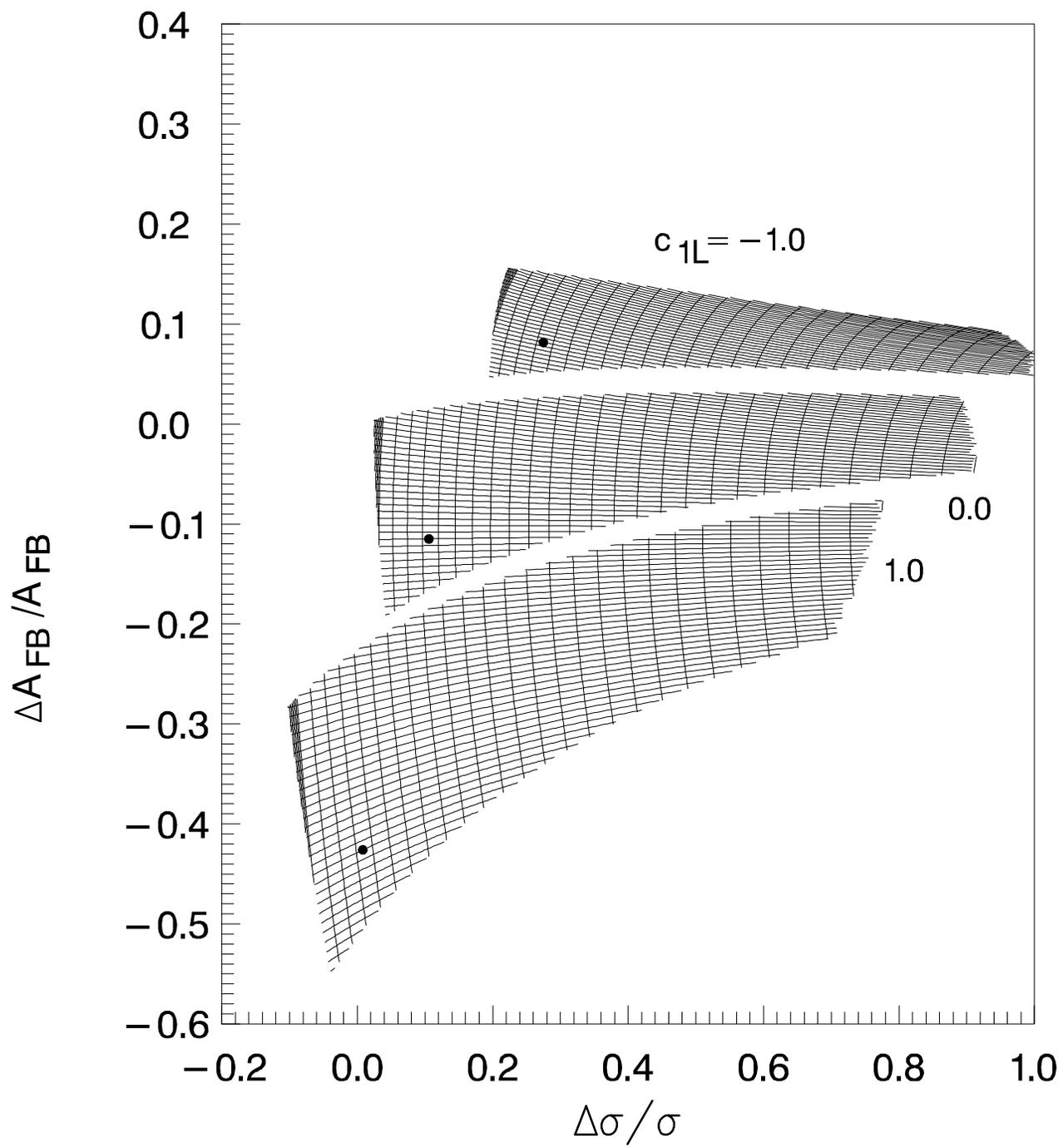
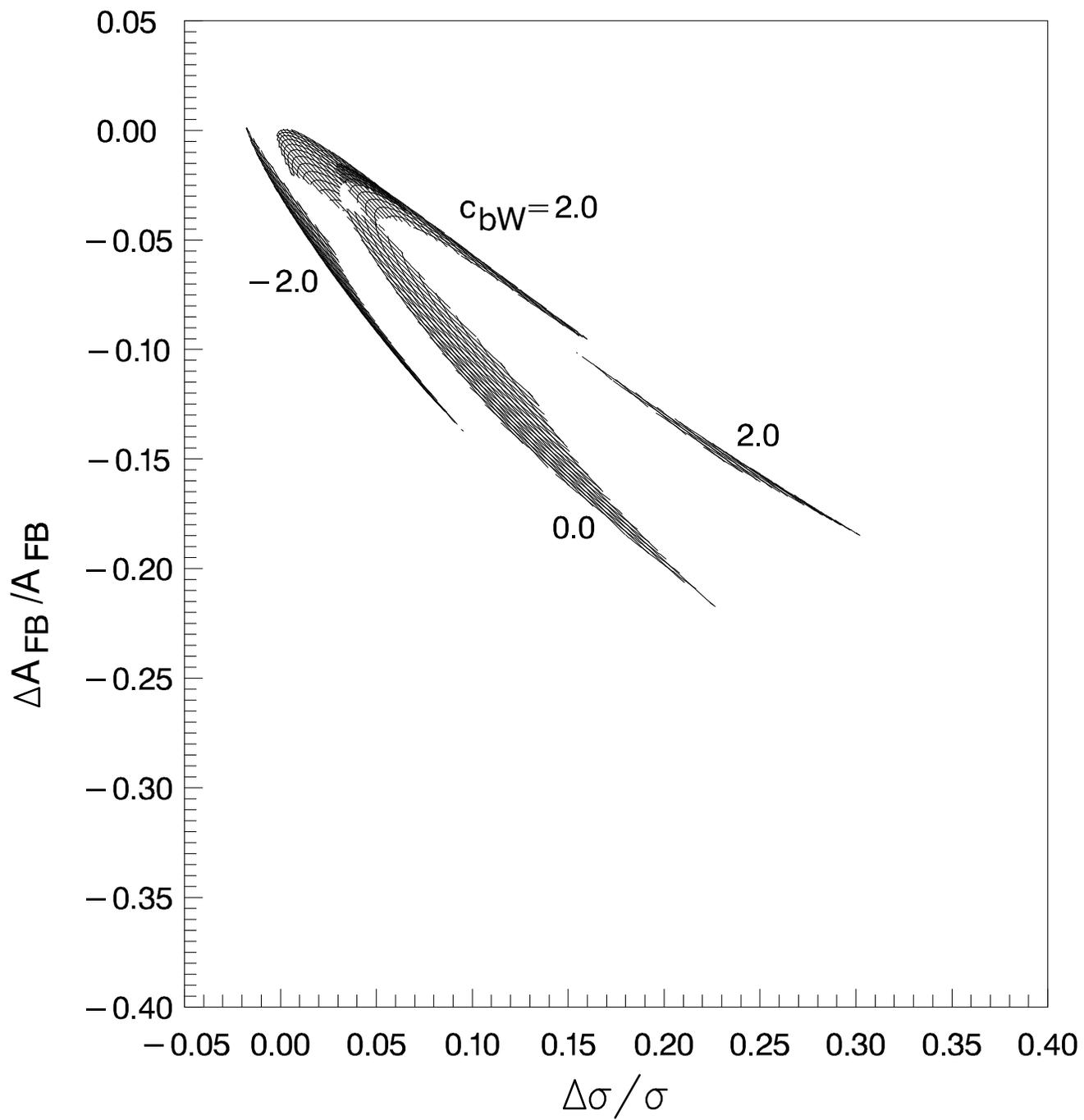
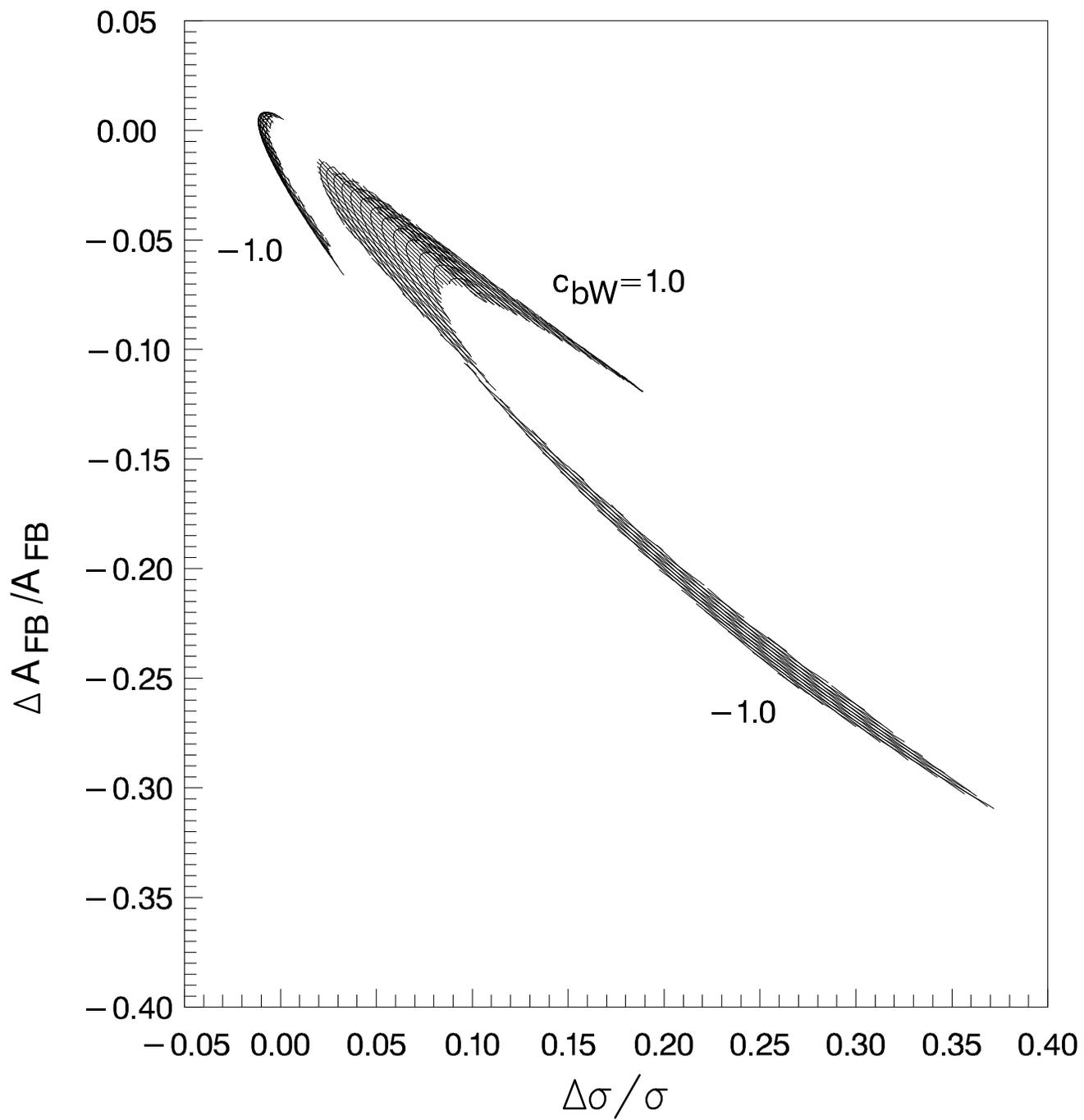
\end{document}